\documentclass[3p,times]{elsarticle}
\usepackage{ecrc}

\volume{00}
\firstpage{1}
\journalname{Nuclear Physics A}

\runauth{J.P.~Lansberg}

\jid{nupha}

\usepackage{amssymb}
\biboptions{sort&compress}

\usepackage[figuresright]{rotating}
\usepackage{amsmath}
\usepackage{color}
\usepackage{graphicx}
\usepackage{float} 
\usepackage{subfig} 
\usepackage{ifpdf}
\ifpdf
\usepackage[pdftex]{hyperref}
\else
\usepackage[hypertex]{hyperref}
\fi

\hypersetup{
  pdftitle={},%
  pdfauthor={},%
  pdfsubject={},%
  pdfkeywords={},%
  pdfstartview={},%
  bookmarksopen=true, breaklinks=true, debug=true, %
  colorlinks=true, linkcolor=red, citecolor=blue, urlcolor=blue
}

\newcommand{\beq}{\begin{eqnarray}}
\newcommand{\eeq}{\end{eqnarray}}
\newcommand{\be}{\begin{eqnarray*}}
\newcommand{\ee}{\end{eqnarray*}}

\newcommand{\eqs}[1]{\begin{equation} \begin{split} #1\end{split} \end{equation} }

\newcommand{\eg}{{\it e.g.}}

\newcommand{\cf}[1]{{Fig.~\ref{#1}}}

\def\lsim{\raise0.3ex\hbox{$<$\kern-0.75em\raise-1.1ex\hbox{$\sim$}}}
\def\gsim{\raise0.3ex\hbox{$>$\kern-0.75em\raise-1.1ex\hbox{$\sim$}}}

\newcommand{\etal}{{\it et al.}}

\def\jpsi   {\mbox{$J/\psi$}}

\def\beq     {\begin{equation}}
\def\eeq     {\end{equation}}

\long\def\symbolfootnote[#1]#2{\begingroup%
  \def\thefootnote{\fnsymbol{footnote}}\footnote[#1]{#2}\endgroup}

\begin{document}

\begin{frontmatter}

\dochead{}

\title{$\Upsilon$ production in $pp$ and $pA$ collisions: from RHIC to the LHC}

\author{J.~P.~Lansberg}
\address{IPNO, Universit{\'e} Paris-Sud, CNRS/IN2P3, 91406 Orsay, France}

\begin{abstract}
I discuss $\Upsilon$ production in $pp$ collisions at RHIC, Tevatron and LHC energies, in particular
the behaviour of the differential cross section in rapidity and the impact of QCD corrections on the 
$P_T$ differential cross section. I also emphasise the very good agreement between the 
parameter-free predictions of the Colour-Singlet Model (CSM) and the first LHC data, especially in the 
region of low transverse momenta, which is the most relevant one for heavy-ion studies. 
I also show that the CSM predicts  $\Upsilon$ cross-section ratios in agreement with the most recent 
LHC data.  I then briefly discuss the nuclear-matter effects on $\Upsilon$ production at RHIC and the LHC in 
$p(d)A$ collisions and, by extension, in $AA$ collisions. I argue that a) the $\Upsilon$ 
break-up probability can be neglected, at RHIC and the LHC, b) gluon shadowing 
--although non-negligible-- 
is not strong enough to describe  forward RHIC data, c) backward RHIC data hints at a gluon EMC effect, 
possibly stronger than the quark one. Outlooks for the LHC $p$Pb run are also presented.
\end{abstract}


\end{frontmatter}

\section{Introduction}
\label{sec:intro}

With the advent of the LHC, the study of $\Upsilon$ production has become more accessible than ever. First 
results~\cite{Khachatryan:2010zg,Aad:2011xv,Aaij:2012ve,Chatrchyan:2011pe},  
in $pp$ and PbPb  collisions, have already been obtained and the $\Upsilon$ production pattern
at the LHC differs from that of the lighter $\psi$'s. $\Upsilon$'s
are thus complementary probes of the QCD dynamics   in $pp$ and PbPb collisions besides the charmonia~\cite{Lansberg:2006dh}. 
It is therefore
important to achieve a good understanding of their production mechanism in the vacuum as well as of
how different the nuclear effects in proton-nucleus collisions are when they act on $\Upsilon$ and on $J/\psi$.

I first discuss $pp$ collisions. I show that the $P_T$-integrated yields obtained at LO in $\alpha_S$ 
and $v$ (the $b$-quark velocity in the $\Upsilon$) agree with the data at different
$\sqrt{s}$ and $y$. In turn, I briefly mention the impact of QCD corrections on the $P_T$ spectrum. In particular, a
comparison with the LHCb data is shown and it demonstrates that the NLO CSM describes very well the $\Upsilon$ yield up to 
5 GeV. The CSM also provides with parameter-free predictions for $\Upsilon$ cross-section 
ratios in agreement with the most recent LHC data. 
Finally, the NNLO leading-$P_T$ contributions seem to be required to account for the data
at larger $P_T$ as the comparison with the NNLO$^\star$ yield shows. In a second section, I discuss the effect 
of (cold) nuclear matter as probed in $\Upsilon$ production 
in $d$Au collisions at RHIC and $p$Pb collisions at the LHC.

\section{$\Upsilon$ production in $pp$ collisions: from RHIC to LHC energies}
\label{sec:pp}

\subsection{Total and differential cross sections}

 I discuss first  the total number of $\Upsilon$ produced in $pp$ collisions as predicted by the CSM at LO in $\alpha_S$ and
irrespective of their transverse momenta. Contrary to what is sometimes claimed in the literature (see \eg~\cite{Cooper:2004qe})
the yield from colour-singlet transitions agrees with the experimental measurements. There is a slight discrepancy at RHIC energy
with the STAR data~\cite{Abelev:2010am}, which is nevertheless less precise\footnote{The number of events is lower, the 3 $\Upsilon$
states are not resolved and the feed-down from $\chi_b$ has never been measured at this energy.} 
than that of the Tevatron~\cite{Acosta:2001gv,Abazov:2005yc} and LHC~\cite{Khachatryan:2010zg,Aad:2011xv,Aaij:2012ve}, 
with which the LO CSM evaluations are in good agreement.

\cf{fig:CSM-xsection} (a) and (b) nicely illustrate  the situation. Both the energy and the 
rapidity dependences  of the $P_T$-integrated cross section are well reproduced by the LO band.
The theory uncertainty at LO is unfortunately large due to the presence of three powers of $\alpha_S$ in the LO cross-section, 
hence one finds a significant renormalisation-scale dependence. The experimental measurements at the 
Tevatron and the LHC are in fact more precise than the theory. Yet, it has to be noted that, at the LHC (\cf{fig:CSM-xsection} (b)), 
the experimental points tend to lie in the lower part of the theory band.

\begin{figure*}[htb!]
\begin{center}\vspace*{-.5cm}
\hspace*{-.1cm}\subfloat[][Energy dependence of the $\Upsilon(1S)$ yield \\ at $y=0$]{
\includegraphics[width=0.36\linewidth]{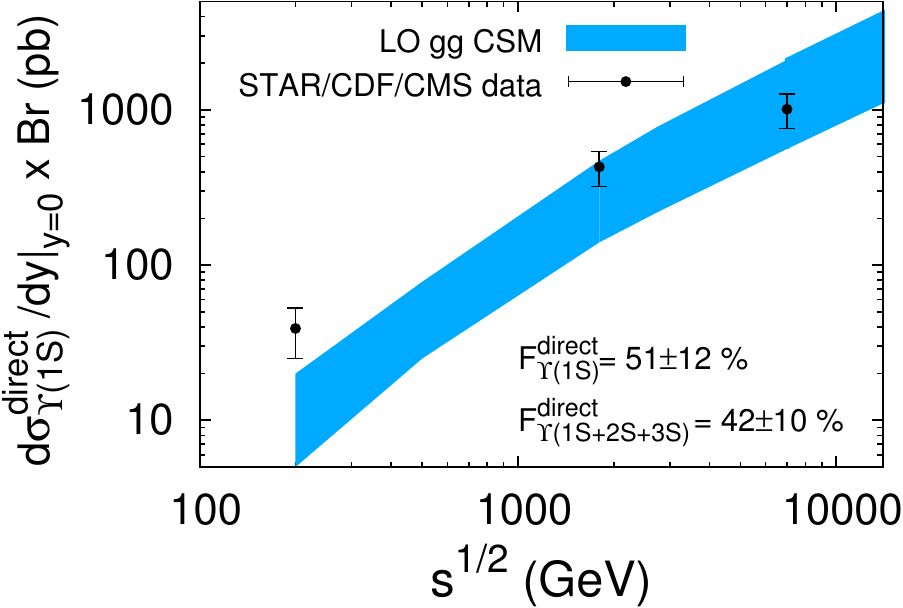}
}
\hspace*{-.1cm}\subfloat[][Rapidity dependence of the $\Upsilon(1S)$ yield \\ at 7 TeV]{
\includegraphics[width=0.36\linewidth]{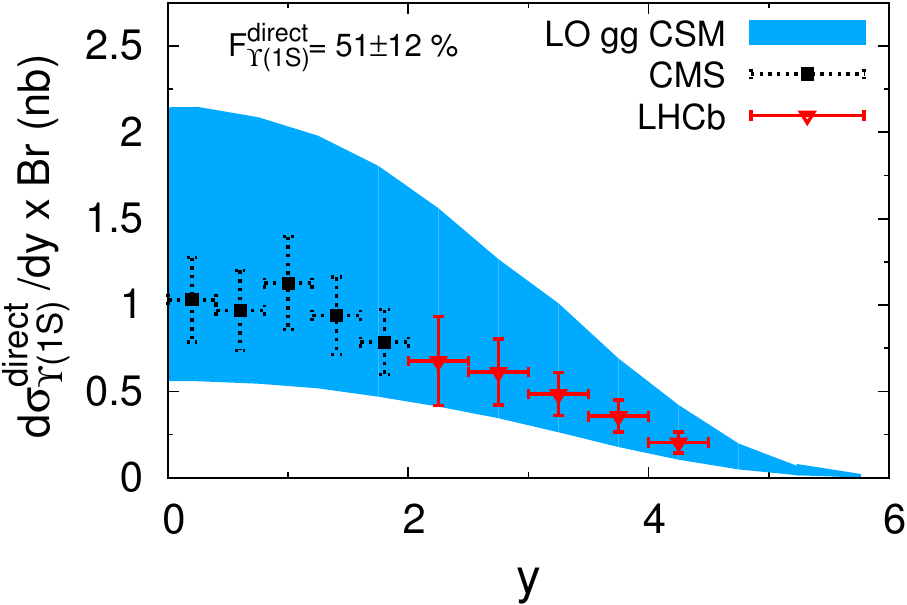}
}
\subfloat[][$P_T$ dependence of the $\Upsilon(3S)$ yield \\ at 7 TeV and 
for $2.0 < y < 4.5$]{
\includegraphics[width=0.28\linewidth]{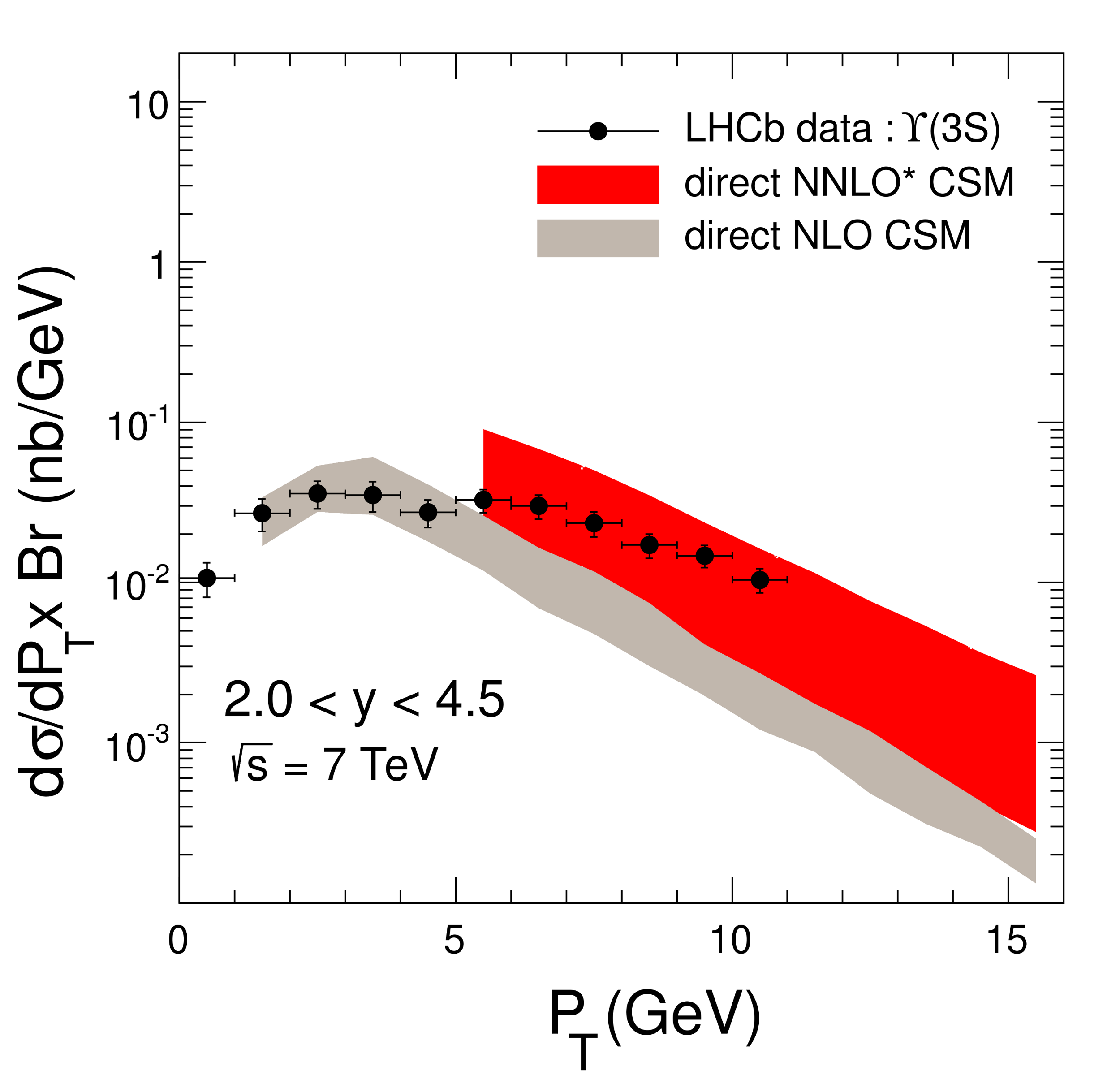}
}
\vspace*{-0.4cm}
\end{center}
\caption{(a) and (b): comparison between the CSM predictions for the direct $\Upsilon(1S)$ yield and various experimental
data~\cite{Abelev:2010am,Khachatryan:2010zg,Aaij:2012ve}  
for the prompt $\Upsilon(1S)$ yield multiplied by $F^{\rm direct}_{\Upsilon(1S)}$~\cite{Affolder:1999wm}
or $F^{\rm direct}_{\Upsilon(1S+2S+3S)}$~\cite{Brodsky:2009cf} for the STAR data. (c): 
comparison between the $\Upsilon(3S)$ LHCb data~\cite{Aaij:2012ve} and the NLO and NNLO$^\star$ 
CSM predictions for the direct yield.}
\label{fig:CSM-xsection}
\end{figure*}

The situation is nevertheless more intricate when the $P_T$ dependence of the yield is concerned~\cite{Lansberg:2008gk}. 
The main reason
is that the leading-$P_T$ contributions to $\Upsilon$ hadroproduction only appear at NNLO in the CSM. For the time being,
only the NLO cross section~\cite{Campbell:2007ws} is fully known along with a partial evaluation of the NNLO yield, dubbed 
NNLO$^\star$~\cite{Artoisenet:2008fc}. 
As expected from the discussion of the $P_T$ integrated yields, the cross section at low $P_T$ is well reproduced
by the NLO yield; it only differs from the LO yield by a harder $P_T$ spectrum. The partial NNLO yield is even harder
and it matches the data at higher $P_T$. Yet, a full NNLO computation is needed before drawing final conclusions.
This is illustrated on \cf{fig:CSM-xsection} (c) by a comparison between the LHCb data for $\Upsilon(3S)$ compared to the NLO and NNLO$^\star$ CSM predictions for the direct yield. The full NLO evaluation --without any ajustable parameter-- 
perfectly matches the LHCb data up to 5 GeV. The comparison is equally good with the $1S$ and $2S$ states~\cite{Aaij:2012ve} 
provided that one subtracts the part of the yield from feed downs. At larger $P_T$, the leading-$P_T$ contributions of the 
NNLO seem to be required to describe the data. 
Overall, this confirms that this sole CS channel contribution seems to be sufficient to convincingly reproduce the total 
yield~\cite{Brodsky:2009cf} {\it as well as} the cross section differential in 
$P_T$~\cite{Lansberg:2008gk} as measured at RHIC~\cite{Abelev:2010am}, the Tevatron~\cite{Acosta:2001gv,Abazov:2005yc} and the
 LHC~\cite{Khachatryan:2010zg,Aad:2011xv,Aaij:2012ve} --once $P_T^{-6}$ (NLO) and $P_T^{-4}$ (NNLO) contributions are included.

\subsection{Cross-section ratios at LO}

Despite the rather large theoretical uncertainties of the CSM predictions, these are free of any adjustable parameter. 
The overall normalisation or the $P_T$ and $y$ dependence cannot be tuned by fitting non-perturbative 
parameters, for instance. In particular, at LO in $v$, ratios of cross sections for {\it direct} $\Upsilon(nS)$ are obtained straightforwardly. These
are in fact simple ratios of Schr\"odinger's wave function at the origin, $\psi^{nS}(0)$. We have:
\eqs{ \frac{\sigma(\hbox{\small direct } \Upsilon(3S))}{\sigma(\hbox{\small direct } \Upsilon(1S))}=\frac{|\psi^{3S}(0)|^2}{|\psi^{1S}(0)|^2}{\sim 0.34}, ~~
 \frac{\sigma\hbox{\small (direct } \Upsilon(2S))}{\sigma(\hbox{\small direct } \Upsilon(1S))}=\frac{|\psi^{2S}(0)|^2}{|\psi^{1S}(0)|^2}{\sim 0.45}.}

As we mentioned, these numbers hold for the {\it direct} yields. From an early CDF study~\cite{Affolder:1999wm} 
at the Tevatron (1.8 TeV), we know that roughly 50$\%$ of the inclusive $\Upsilon(1S)$'s are directly produced. 
We will make the reasonable assumption that a similar fraction holds at the LHC, with the drawback that  CDF did 
not measure  low-$P_T$ $\chi_b$'s. From the recent CMS measurement~\cite{Khachatryan:2010zg} 
($\sigma(\Upsilon(1S) (|y|< 2))Br_{\ell\ell} \simeq 7.4$ nb), 
we can thus obtain an evaluation of the direct $\Upsilon(1S)$ yield:
$ \sigma(\hbox{\small direct } \Upsilon(1S))\sim 150$ nb.
In turn, it is straightforward to get what is expected from the CSM for the  direct $\Upsilon(3S)$ yield at 7 TeV:  
$0.34 \times 150$~nb~$\sim 50$~nb. This is surprisingly close to the value 
measured\footnote{$\sigma(\Upsilon(3S)(|y|< 2)) Br_{\ell\ell} \simeq 1.0$ nb 
$\overset{{100\% direct}}{\longrightarrow} \sigma(\hbox{\small direct } \Upsilon(3S))\sim$ {45 nb}} by 
CMS~\cite{Khachatryan:2010zg}, 45 nb, assuming that 100$\%$ of the 
$\Upsilon(3S)$ are directly produced. The latter assumption was perfectly sound until recently. 
However, ATLAS has made the first observation~\cite{Aad:2011ih} of a candidate
for the $\chi_b(3P)$ which is likely to decay into $\Upsilon(3S)$. The $\Upsilon(3S)$ yield may not be 100$\%$ direct.

From the above arguments, one would not expect any $P_T$ dependence of these cross-section ratios. Two effects 
should nevertheless be kept in mind. 
At low $P_T$, the $P_T$ dependence of the cross section is known to be affected by mass effects. Indeed, it does not follow a simple
power-law falloff in $P_T$. We should be aware that the ratio predicted above hold in the limit where $M^{\Upsilon(nS)}_{\rm NRQCD}=2 m_b$
for all the states. Such an approximation is not ideal for the $3S$ states for instance.
Another effect comes from a simple kinematical effect of the feed-down: the transverse momentum of a daughter particle
 is always smaller than that of the parent (the excited state here). Quantitatively, we can write the approximate relation:
$P_T^{\rm daughter} \! \! \sim ({M^{\rm daughter}}/{M^{\rm mother}}) \times P_T^{\rm mother}$.

\begin{figure*}[htb!]
\begin{center}
%
\includegraphics[width=0.35\linewidth]{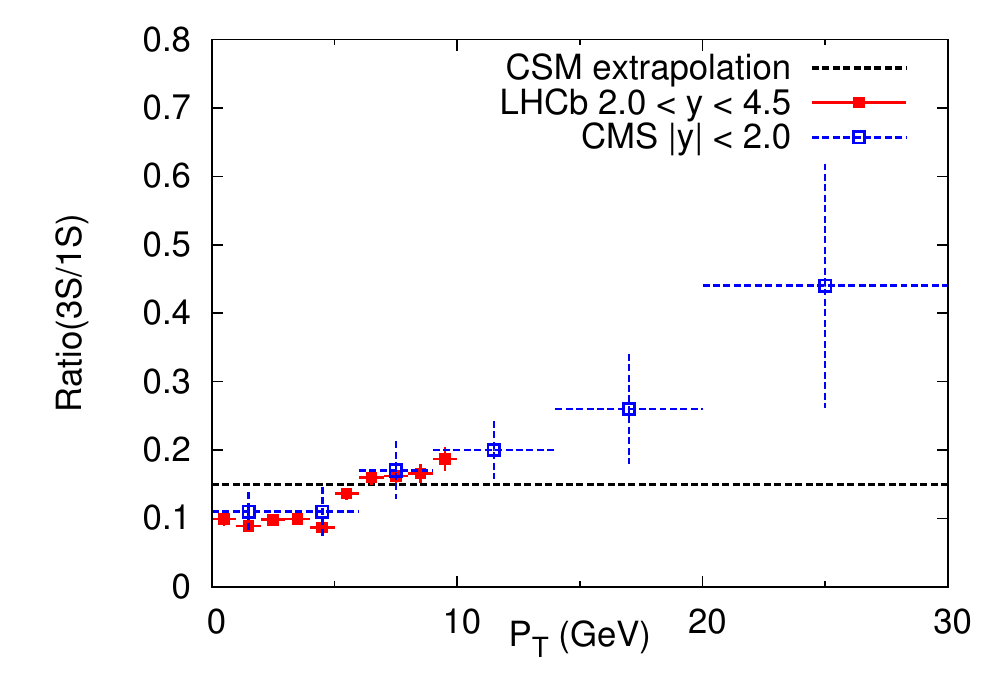}
%
\includegraphics[width=0.35\linewidth]{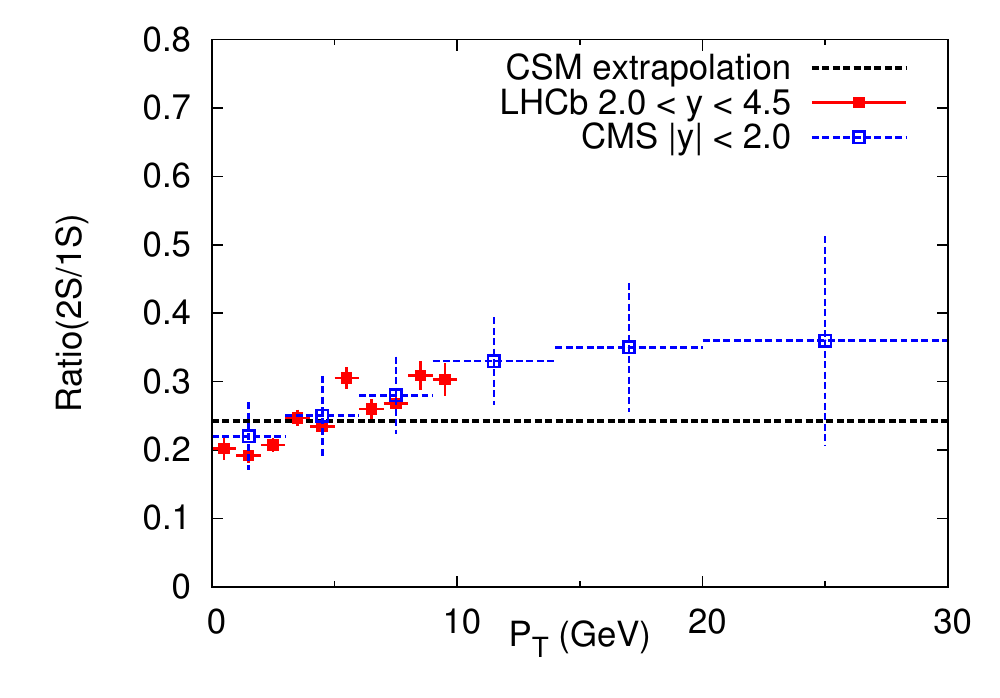}
\end{center}
\vspace*{-.8cm}
\caption{$\Upsilon$ cross-section ratios ``$3S/1S$'' and ``$2S/1S$'' as measured by LHCb~\cite{Aaij:2012ve} and CMS~\cite{Khachatryan:2010zg} and as predicted by the CSM without the kinematical effect mentioned in the text. The systematic experimental uncertainties from the unknown polarisation are not shown.}\vspace*{-0.4cm}
\label{fig:xsection-ratios}
\end{figure*}

Such a rescaling of the $P_T$ spectrum does not matter if $\frac{d\sigma}{dP_T} \propto P_T^{-n}$ with $n$ fixed  
--we would always keep  the cross-section ratio independent of $P_T$ (with a smaller feed-down for higher $n$, though). 
However, if $n$ changes, which is especially true at low $P_T$, this induces a $P_T$ dependence of the feed-down, even 
if both the direct production cross sections of the lower-lying state and of the excited state have the same $P_T$ 
dependence. The feed-down fraction is expected to increase when the $P_T$ dependence is mild
and to decrease when it is steep. From the usual shape of $\frac{d\sigma}{dP_T}$, one expects a decrease of the feed-down 
with $P_T$. In turn, the cross-section ratio, $\Upsilon(mS)/\Upsilon(nS)$ with $m>n$ would increase with $P_T$ until 
the differential cross section gets the behaviour of a power law ($P_T^{-n}$). A detailed Monte Carlo simulation is required 
to quantify this increase. Yet, it is instructive to keep in mind that $(M_{\Upsilon(3S)}/M_{\Upsilon(1S)})^6\simeq 1.7$. 
Such kinematical effects up to 50$\%$ would not be surprising. 
This provides a reasonable explanation of the ratio increase observed by CMS~\cite{Khachatryan:2010zg} and 
LHCb~\cite{Aaij:2012ve} on Figs.~\ref{fig:xsection-ratios}.

\section{$\Upsilon$ production in $pA$ collisions: from RHIC to LHC energies}
\label{sec:pA}

\subsection{$d$Au collisions at RHIC}

As we discussed in~\cite{Ferreiro:2011xy}, the survival probability of the $\Upsilon$ (or its pre-resonant state) in the 
nuclear matter is significantly larger than for $J/\psi$ -- or conversely its ``absorption'' is small. 
This is due to its smaller size. On the contrary, the impact of (anti)shadowing is not necessarily small. 
When compared to $J/\psi$ results (see \eg~\cite{Ferreiro:2008qj}), 
two effects should be kept in mind. First, the energy scale of the scattering entering the 
nuclear-PDF evaluation ($Q$ or $\mu_F$) is expected to be three times larger for the $\Upsilon$ than for the $J/\psi$. 
A priori, we do not expect any saturation effect of the gluon densities~\cite{Ferreiro:2011xy}. Second, 
and more importantly, the average 
momentum fractions of the partons in both colliding particles are three times larger in the $\Upsilon$ case compared 
to that of $J/\psi$'s for a fixed quarkonium rapidity. This explains, for instance, 
why anti-shadowing could show up at a given rapidity for $\Upsilon$ and not for $J/\psi$.

At RHIC, $\Upsilon$ production allows one to probe the gluon densities in a wide momentum-fraction range: 
at forward $y$, it probes the shadowing region; at mid $y$, the anti-shadowing region; and, at backward $y$, the 
EMC region. The latter is pretty much unknown for gluons and we emphasised 
in~\cite{Ferreiro:2011xy} that $\Upsilon$  production in $d$Au collisions at RHIC energies can be an invaluable probe 
of the unexplored large-$x$ dynamics of gluons in bound nucleon inside heavy nucleus (see also~\cite{Lansberg:2012kf}). 
As a matter of fact, the  slight $\Upsilon$ suppression at backward $y$ observed by PHENIX~\cite{Lee:2010zzr}  nearly rules out a gluon 
excess from Fermi motion for $x$'s close to 0.3 as predicted in~\cite{Merabet:1993du} and it
may even be the first hint of gluon EMC suppression.

At mid $y$, an update of the STAR data~\cite{Reed:2010zzb} is awaited to confirm the absence 
of a significant anti-shadowing. On the contrary, the forward PHENIX data~\cite{Lee:2010zzr} are already precise 
enough to indicate that an ingredient is missing in the existing analyses since the conventional
shadowing is far from being enough~\cite{Ferreiro:2011xy} to explain the large suppression of the $\Upsilon$ yield.
Fractional parton energy loss for forward-angle quarkonium production might explain this anomalous suppression, 
along the same lines as the analysis of Ref.~\cite{Arleo:2010rb} 
for the \jpsi\ production in proton-nucleus collisions.

\subsection{Outlooks for $p$Pb collisions at the LHC}
Taking into account the rapidity shift between the lab system and the c.m.s. of the colliding particles due to the
imbalance in their energies at the LHC at $\sqrt{s_{NN}}=5$~TeV, one expects an excess of $\Upsilon$ 
between 5 and 15$\%$ due to the anti-shadowing in Pb$p$ collisions in the acceptance of the ALICE detector. In $p$Pb collisions, 
for which the LHCb detector will also take data, one expects a suppression ranging from 20 to 25$\%$ in the forward 
rapidity region.

Such numbers are not to be overlooked: remember that in PbPb collisions, 
such nuclear effects act roughly speaking quadratically compared to in $p$Pb collisions !
Cold nuclear effects on the $\Upsilon$ yield can thus be up to 20-30$\%$ in PbPb collisions. This calls for a 
detailed $\Upsilon$ analysis in the forthcoming $p$Pb/Pb$p$ runs at the LHC.

Along these lines, let us keep in mind that, at RHIC energies, the suppression of $\Upsilon$
in $d$Au collisions is of the same size as that of $J/\psi$.  Novel effects such as fractional energy loss might be 
at work  and need be subtracted for a proper
analysis of the $\Upsilon$ behaviour in the deconfined matter created in central heavy-ion collisions at the LHC.\vspace*{0.2cm}

{\bf Acknowledgements.} 
I thank P. Artoisenet, S.J. Brodsky, J. Campbell, E.G. Ferreiro, F. Fleuret, F. Maltoni, N. Matagne, A. Rakotozafindrabe 
and F. Tramontano for
our fruitful collaborations on the topics presented here.


\bibliographystyle{elsarticle-num}


\end{document}